\documentclass[12pt,aps,a4paper,nofootinbib,floatfix,reprint]{revtex4-2}
\usepackage[utf8x]{inputenc}
\usepackage[T1]{fontenc}
\usepackage{tikz}
\usetikzlibrary{quantikz}
\usepackage{mathptmx} % Use Times Font
\usepackage{mathtools}
\usepackage{amsmath}
\usepackage{amssymb}
\usepackage{subcaption}
\usepackage{braket}
\usepackage[export]{adjustbox}

\usepackage{multirow}
\usepackage{graphicx} % Required for including pictures
\usepackage{hyperref}
\usepackage[a4paper, lmargin=0.1666\paperwidth, rmargin=0.1666\paperwidth, tmargin=0.1111\paperheight, bmargin=0.1111\paperheight]{geometry} %margins
\usepackage{caption}
\begin{document}
\title{Quantum computation of nuclear observables involving linear combination of unitary operators}
\author{Pooja Siwach}\email{psiwach@physics.wisc.edu}
\affiliation{Department of Physics, University of Wisconsin--Madison, Madison, Wisconsin 53706, USA}
\author{P. Arumugam}\email{arumugam@ph.iitr.ac.in}
\affiliation{Centre for Photonics and Quantum Communication Technology, and Department of Physics, Indian Institute of Technology Roorkee, Roorkee 247667, Uttarakhand, India}
\date{\today}

\begin{abstract}
    We present the quantum computation of nuclear observables where the operators of interest are first decomposed in terms of the linear combination of unitaries. Then we utilise the Hadamard test and the linear combination of unitaries (LCU) based methods to compute the expectation values. We apply these methods to calculate the electric quadrupole moment of deuteron. The results are compared for the Jordan-Wigner transformation and Gray code encoding. We discuss the versatility of our approach that can be utilized in general to calculate several observables on a quantum computer.
\end{abstract}

\maketitle

%###############
\section{Introduction}
Quantum computing has become a reality now and the importance to build various blocks of imminent applications is immense. One among such applications is in the field of many-body theory where the computational challenges are galore.  Considerable progress in this direction is evident in quantum chemistry~\cite{Cao:2019,Bauer:2020, Mcardle:2020} and several areas of physics~\cite{Alexandru:2019, Macridin:2018, Lamm:2019}.  In nuclear physics, such attempts have gained momentum recently~\cite{Dumitrescu:2018,Lu:2019,Shehab:2019,du2020,Roggero:2019,Roggero:2020,Roggero1:2020,Denis:2020,Pooja:qc,Siwach:PRA2021,cervia2021,Denis:2020,Andres:2022,Hlatshwayo:2022}.  The present work is aimed to augment such efforts by providing solutions for calculating expectation value of operators on the quantum computer by utilizing the wave functions obtained through quantum simulations.  

In this work, we propose mainly two methods to compute the expectation values of non-unitary operators. First, we decompose the non-unitary operators in terms of unitary ones by expressing the operators in the second quantization form. The expectation value of these linear combination of unitaries (LCU) can be easily computed on the quantum computer using Hadamard test as used in VQE algorithm. Second, we implement the LCU method~\cite{childs2012hamiltonian,Childs:2017} to calculate the operation of non-unitary operation on the wave function. This technique has been proposed to prepare the excited state on a quantum computer for a nuclear system.~\cite{Roggero:2020}. Here, we extend it to compute the expectation value of non-unitary operators. A SWAP test and destructive SWAP test~\cite{swap:2013} are used to calculate the overlap of resulting state with the original wave function which gives the required expectation value. The detailed formalism and techniques are given in the subsequent sections.

As an illustrative application, we employ these algorithms to calculate the electric quadrupole moment of deuteron. First, the binding energy of deuteron is calculated with the VQE algorithm, and then the resulting ground state wave function is utilized to calculate the quadrupole moment.  Though the earlier works~\cite{Dumitrescu:2018, Pooja:qc} could successfully deploy quantum simulations and demonstrate in the case of deuteron, the accuracy of the resulting binding energy is not up to the mark.  This is expected due to the increase in errors while increasing the basis size which results in larger circuit depth and more number of gates. This error is further augmented by the limitations in the number of qubits available in the present quantum computers.  With such restricted hardware, we need to work with limited basis size and results from classical computing are used to benchmark the quantum simulations.

% \vspace{-0.02cm}
This paper is arranged as follows. In Section~\ref{sec:methods}, we detail both the methods implemented in present work to calculate the expectation values of non-unitary operators. In section~\ref{sec:nucl}, we provide the detailed expressions for the nuclear operators and quantum circuits for calculating their expectation values with both the methods. A few simple working examples are also given to help the reader to follow our implementations. The results are given in Section~\ref{sec:result} and this work is concluded in Section~\ref{sec:conc}. More details for the algorithms are given in Appendices~\ref{app:swap} and~\ref{app:vp}.

\section{Quantum computation of expectation values}\label{sec:methods}
\subsection{Method based on linear combination of unitary operations}
\begin{figure}[h!]
% \begin{tikzpicture}
% \node[scale=0.7] {
% \begin{quantikz}
% \ket{0}&\qwbundle{}&\gate{V_{P}}&\gate[wires=2][2cm]{V_{S}}&\gate{V_{P}^{\dagger}}&\qw\\
% \ket{\psi}&\qwbundle{}&\qw& &\qw&\qw
% \end{quantikz}
% };
% \end{tikzpicture}
\includegraphics[width = 0.9\linewidth]{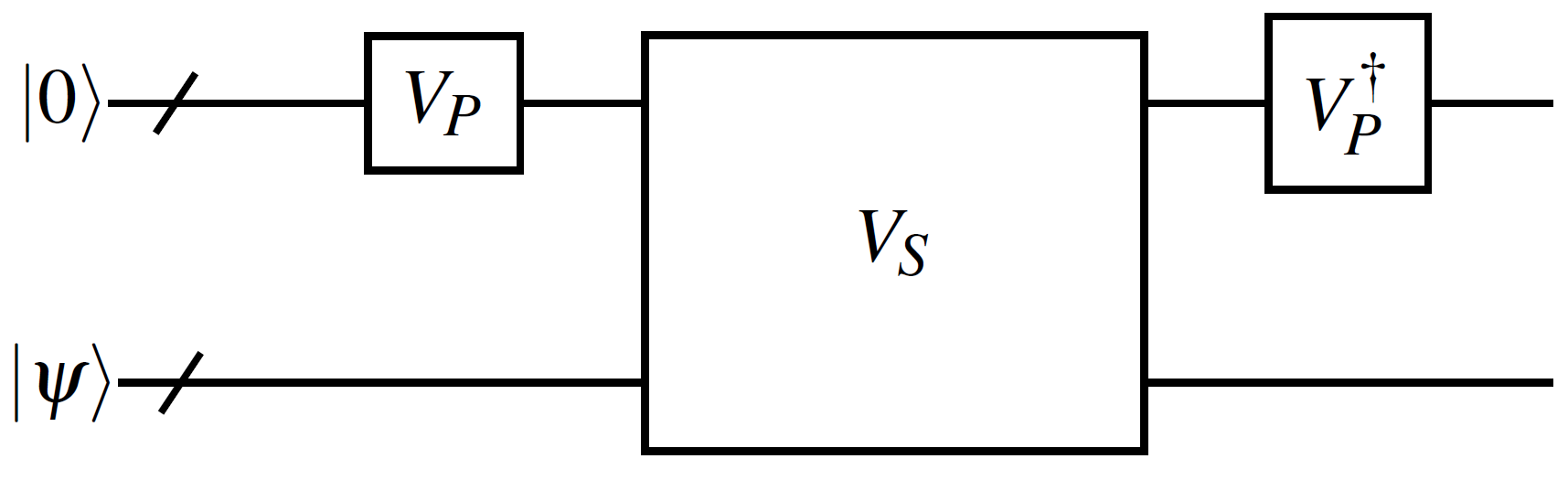}
\caption{The quantum circuit to calculate $\mathcal{O}\ket{\psi}$ where $\mathcal{O}$ is represented as a linear combination of unitaries (LCU) through $V_P$ and $V_S$.}\label{fig:w}
\end{figure}

The method based on the linear combination of unitaries (LCU) was first proposed in Ref.~\cite{childs2012hamiltonian} for Hamiltonian simulations~\cite{Childs:2017}. Given an operator $\mathcal{O}=\sum_{i=0}^{k-1}\beta_{i}U_{i}$ that is a linear combination of unitaries $U_{i}$ (or a block of unitaries) with $\beta_{i}>0$ for all $i$, and an operator $V_{P}$ that satisfies
\begin{equation}\label{eq:vp1}
    V_{P}\ket{0}^{\otimes n_{a}}=\frac{1}{\sqrt{\Lambda}}\sum_{i}\sqrt{\beta_{i}}\ket{i},
\end{equation}
where $\Lambda=\sum_{i}\beta_{i}$ and $\ket{i}$ is the $i^{\rm th}$ component of the state, we can define an operator $W=V_{P}^{\dagger}V_{S}V_{P}$ that satisfies
\begin{equation}\label{eq:w}
    W\ket{0}^{\otimes n_{a}}\ket{\psi}=\frac{1}{\Lambda}\ket{0}^{\otimes n_{a}}\mathcal{O}\ket{\psi}+\ket{\Psi^{\perp}}.
\end{equation}
 The quantum circuit corresponding to $W$ is shown in \figurename~\ref{fig:w}. Here $V_{s}=\sum_{i=0}^{L}\ket{i}\bra{i}\otimes U_{i}$ is sometimes called the \textit{select} operator. The $\ket{\Psi^{\perp}}$ is orthogonal to the $\ket{0}^{n_{a}}$ of ancilla register, i.e., $\ket{0}\bra{0}\otimes \ket{\Psi^{\perp}}=0$. $n_{a}=\left\lceil\log_{2} k\right\rceil$ is the number of ancillary qubits. Hence, we obtain the desired state $\mathcal{O}\ket{\psi}$ (up to normalization with normalization constant $\eta=||\mathcal{O}\ket{\psi}||$) when the ancilla register is in $\ket{0}$ state.
 
\begin{figure*}%[h!]
\includegraphics[width=0.9\linewidth]{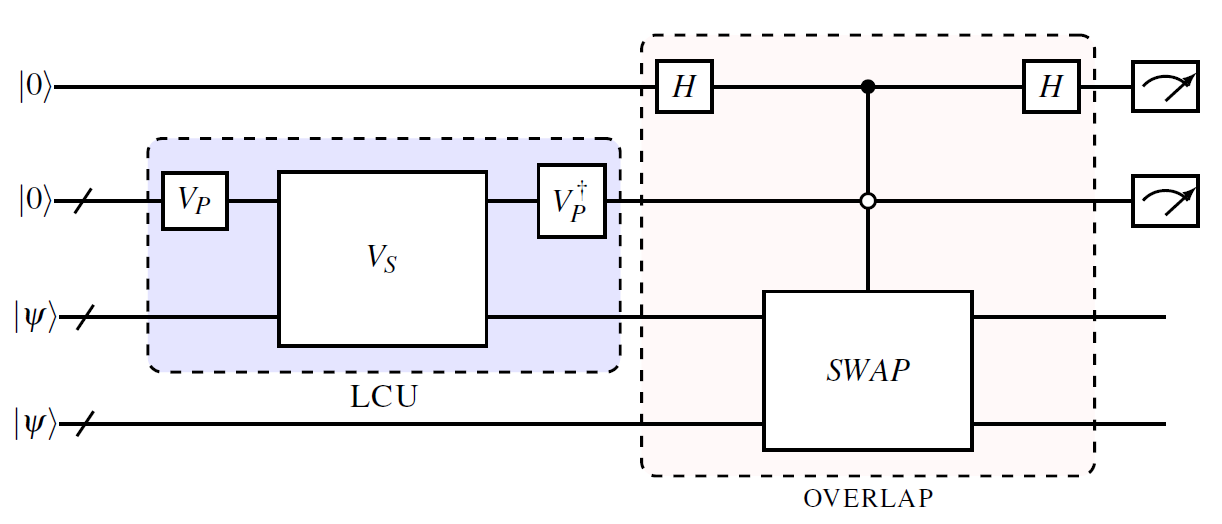}
\includegraphics[width=0.9\linewidth]{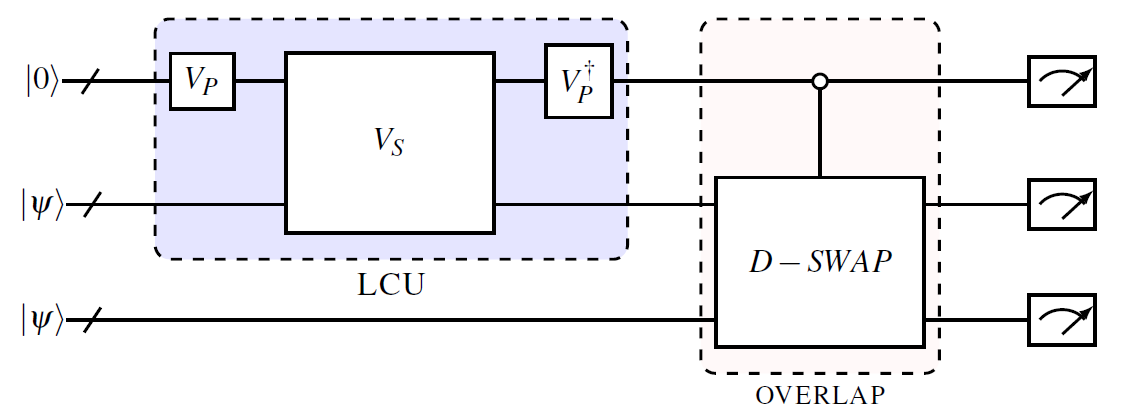}
\caption{The quantum circuit to calculate the expectation value of an opearator by combining the LCU method [in blue blocks to calculate $\mathcal{O}\ket{\psi}$] and the SWAP [top] and destructive SWAP (D-swap) [bottom] test [in red blocks to calculate $\bra{\psi}\mathcal{O}\ket{\psi}$].}\label{fig:wcomp}
\end{figure*}

 To further compute the expectation value $\bra{\psi}\mathcal{O}\ket{\psi}$, we need to calculate the overlap of $\ket{\psi}$ and $\mathcal{O}\ket{\psi}$. This operation can be performed using the SWAP test or the destructive SWAP test. The details on these algorithms can be found in \appendixautorefname~\ref{app:swap}. The complete circuit is shown in \figurename~\ref{fig:wcomp}.
 
%$$$$$$$$$$$$$$$$$$$$$$$$$$$$$$$$$$$$$$$$$$$$$$$$$$$$$$$$$$$$$$$$$$$$$$$$$$$$$$$$$$$$$$$$$$$$
\subsection{Method based on Hadamard test}\label{sec:htest}
\begin{figure}[h!]
\includegraphics[width=0.9\linewidth]{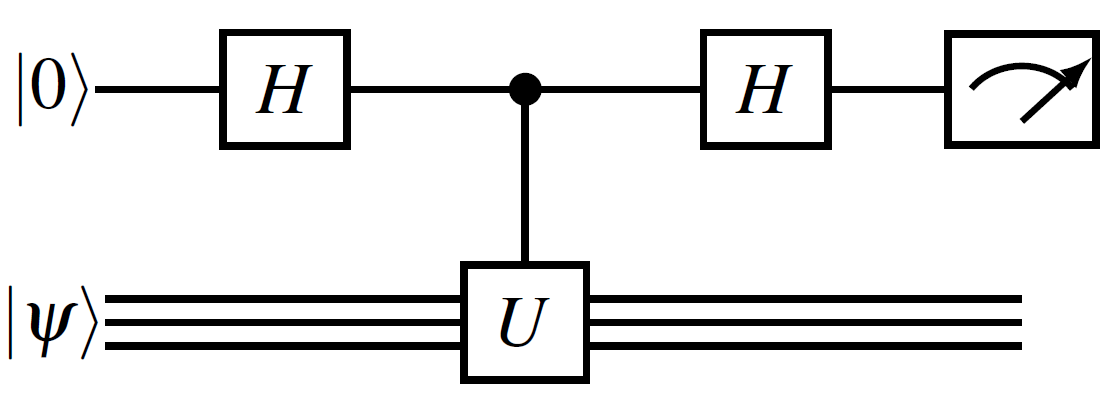}
\caption{The quantum circuit corresponding to Hadamard test (H-test) for calculating Re$(\bra{\psi}U\ket{\psi})$.}\label{fig:had1}
\end{figure}

Another way to compute the expectation value $\bra{\psi}\mathcal{O}\ket{\psi}$ is to do Hadamard test (H-test) for each unitary component separately. The Hadamard test on $U_{i}$ gives us $\bra{\psi}U_{i}\ket{\psi}$. The circuit corresponding to Hadamard test is shown in \figurename~\ref{fig:had1}.

Measuring the ancillary qubit gives us the probability of it being $0 ~(p(0))$ and $1 ~(p(1))$. We get the expectation value Re$(\bra{\psi}U\ket{\psi})$ as the difference between $p(0)$ and $p(1)$. To calculate the Img$(\bra{\psi}U\ket{\psi})$, one can use a $\sqrt{Z}$ gate on ancillary qubit after the controlled unitary operation.

Though the H-test requires lesser number of gates and qubits as compared to the LCU method, it is useful only when we need to compute Re$(\bra{\psi}U\ket{\psi})$ and not Re$(\bra{\psi'}U\ket{\psi})$. In that sense, LCU method is more versatile.
%$$$$$$$$$$$$$$$$$$$$$$$$$$$$$$$$$$$$$$$$$$$$$$$$$$$$$$$$$$$$$$$$$$$$$$$$$$$$$$$$$$$$$$$$$$$$
\section{Expectation values of nuclear operators}\label{sec:nucl}
% $$$$$$$$$$$$$$$$$$$$$ Intro $$$$$$$$$$$$$$$$$$$$$$$$$$$$$$$$$$$$$$$$$$$$$$$$$$$$
We start with calculating the ground state energy of the Hamiltonian using the variational quantum eigensolver (VQE). Similar to Ref.~\cite{Pooja:qc}, we consider as an example, the ground state of deuteron using the REID68 potential~\cite{REID1968} which can account for the admixture of orbital angular momentum $l=0$ and $2$ states. As presented in Ref.~\cite{Pooja:qc}, the VQE fails for this particular potential to converge to the ground state. This issue can be resolved by defining the operators using \texttt{WeightedPauliOperator} instead of \texttt{MatrixOp} module of qiskit. We suspect some bug in the latter module. \footnote{Simulation results with QASM simulator are different when we define operators using \texttt{MatrixOp} and \texttt{WeightedPauliOperator}. The latter gives reasonable results for all operators we tested. For example, the eigenvalue of $H=0.5YY$ is correct for operators defined in both ways, whereas the eigenvalue of $H=-0.5YY$ is not reasonable when we define it using \texttt{MatrixOp}.} Overcoming this issue, we proceed with the ground state obtained from VQE to compute the other properties of deuteron, namely the electric quadrupole moment. Note that the potential derived from EFT used in most of the similar studies~\cite{Dumitrescu:2018, Lu:2019,dimatteo2021} cannot account for the admixture of orbital angular momentum $l=0$ and $2$ states. Therefore, we consider the REID68 potential to calculate the electric quadrupole moment of deuteron.

First, we rewrite the operator of interest $\mathcal{O}$ in the second quantization form such that
\begin{equation}\label{eq:Osec}
    \hat{\mathcal{O}}=\sum_{i,j}\bra{\phi_{i}}\mathcal{O}\ket{\phi_{j}}a_{i}^{\dagger}a_{j},
\end{equation}
where $\ket{\phi_{i}}$ are the harmonic oscillator basis states which are utilised to calculate the eigenvalues and eigenfunctions of the Hamiltonian. We can express the raising/lowering operators $a^{\dagger}/a$ in terms of Pauli spin matrices with the Gray code or direct encoding (Jordan-Wigner transformation)\footnote{We use direct encoding and Jordan-Wigner transformation interchangeably.}. 

We apply our approach to calculate the electric quadrupole moment of deuteron with the operator given by
\begin{equation}\label{eq:O}
    \hat{Q}=\sqrt{\frac{\pi}{5}}r^{2}Y_{20}(\theta,\phi),
\end{equation}
where $r$ is the relative distance between the proton and the neutron, and $Y$ are the spherical harmonics.  The corresponding experimental value is $0.282(1)$ fm$^{2}$.  We calculate the expectation value $Q$ with the deuteron wave function. With the REID68 potential, the full classical calculations with numerical solutions for the coupled differential equations yield the binding energy $-2.2247$ MeV and the quadrupole moment $0.2762$ fm$^{2}$. These values can be reproduced with the solutions obtained by diagonalising the Hamiltonian in harmonic oscillator basis with a sufficient basis size. However, with restricted basis size, the results are approximate and are taken as reference for the quantum simulations.

Applying Eq.~\eqref{eq:Osec} for $\hat{Q}$ (Eq.~\eqref{eq:O}), the operators with the basis sizes\footnote{Here, the basis is $\ket{n,l}$, where $n$ represents the radial quantum number of harmonic oscillator basis and $l$ represents the orbital quantum number. $l$ is always $0$ and $2$ in present case. Therefore, in case of basis size $2$, we have $n=0$ and $l=0,2$.} $2$ and $4$ are given for the cases of Gray code (GC) and Jordan-Wigner (JW) as follows.

GC:
\begin{eqnarray}
  Q_{2}&=&-0.18144I+0.18144Z_{0}\nonumber\\
  &~&+0.28394X_{0}.\label{eq:q2gc}\\
  Q_{4}&=&-0.2332825I-0.0518425Z_{0}\nonumber\\
  &~&+0.0518425Z_{1}+0.2332825Z_{0}Z_{1}\nonumber\\
  &~&+0.358835X_{0}-0.074895X_{0}Z_{1}\nonumber\\
  &~&-0.23184X_{1}+0.23184Z_{0}X_{1}\nonumber\\
  &~&+0.096985(X_{0}X_{1}+Y_{0}Y_{1}).\label{eq:q4gc}
\end{eqnarray}

JW:
\begin{eqnarray}
    Q_{2}&=&-0.18144I+0.18144Z_{1}\nonumber\\
    &~&+0.14197(X_{0}X_{1}+Y_{0}Y_{1}).\label{eq:q2jw}\\
    Q_{4}&=&Q_{2}-0.285125I+0.285125Z_{3}\nonumber\\
    &~&-0.23184(X_{1}X_{2}+Y_{1}Y_{2})\nonumber\\
    &~&+0.096985(X_{1}X_{3}+Y_{1}Y_{3})\nonumber\\
    &~&+0.2236068(X_{2}X_{3}+Y_{2}Y_{3}).\label{eq:q4jw}
\end{eqnarray}
As we can see, the operators $Q_{2}$ and $Q_{4}$ are not unitary. Therefore, to calculate the expectation values of these operators on a quantum computer, we utilise the techniques given in section~\ref{sec:methods}. As learnt from our previous work~\cite{Pooja:qc}, the Bravyi-Kitaev encoding~\cite{BRAVYI2002, BK:2012} does not provide much advantage in terms of efficiency (compared to the JW case), and hence dropped in this study.
% $$$$$$$$$$$$$$$$$$$$$ #### $$$$$$$$$$$$$$$$$$$$$$$$$$$$$$$$$$$$$$$$$$$$$$$$$$$$
\subsection{For basis $\ket{n=0; l=0,2}$}
From the classical calculations, in the harmonic oscillator (HO) basis $\ket{n,l}$, the energy obtained for $n=0$ and $l=0,2$ is $E=67.9476$ MeV~\cite{Pooja:qc} with the corresponding eigenstate
\begin{eqnarray}\label{eq:gc2psi}
    \ket{\psi_{0}}&=&0.2759\ket{0,0}+0.9611\ket{0,2}\\
    &=&R_{Y}(-3.700868)\ket{0},
\end{eqnarray}
where $R_{Y}(-3.700868)$ is the rotation gate along $Y$-axis with angle $-3.700868$, leading to the quadrupole moment
\begin{equation}
    Q=\bra{\psi_{0}}\hat{Q}\ket{\psi_{0}}=-0.1846.
\end{equation}
To construct the quantum circuit to perform the above calculations, we utilise the operators for $\hat{Q}$ in the GC and  JW given in Eqs.~\eqref{eq:q2gc} and~\eqref{eq:q2jw}, respectively, along with the methods presented in Section~\ref{sec:methods}.
\subsubsection{With LCU method}
To elucidate the LCU method, we provide the details for the simplest case, i.e., basis size $2$ with the GC [Eq.~\eqref{eq:q2gc}]
\begin{equation}
    \hat{Q}=-0.18144I+0.18144Z_{0}+0.28394X_{0}.
\end{equation}
To compute the $Q$, we can ignore the first term and define,
\begin{equation}\label{eq:q1}
    \hat{Q'}=0.18144Z_{0}+0.28394X_{0}
\end{equation}
for which we get the results from classical computing as
\begin{equation}
    Q'=\bra{\psi}\hat{Q'}\ket{\psi}=-0.00316
\end{equation}
\begin{equation}
    Q=-0.18144+Q' = -0.1846.
\end{equation}

\begin{figure*}
\centering
\includegraphics[width=0.9\linewidth]{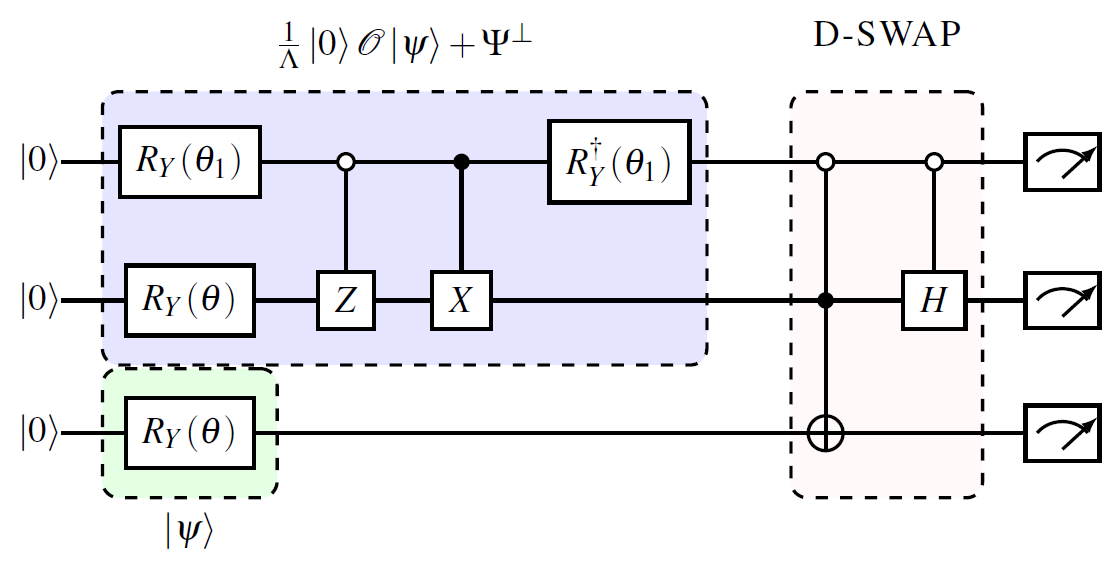}
\caption{The quantum circuit corresponding to LCU method with destructive SWAP (D-SWAP) test for getting expectation value of operator given in Eq.~\eqref{eq:q2gc}. Here $\theta_{1}=1.79286$ and $\theta=-3.700868$ (for the state obtained with classical calculations). The result can be interpreted as $\sqrt{1-2p_{23}(11)_{|p_{1}(0)}}\Lambda$ where $\Lambda=0.46538$, and $p_{ij}(i'j')$ represents the probability of $i^{th}$ qubit being in $i'\in\{0,1\}$ state (similarly for $j^{th}$ qubit being in $j'\in\{0,1\}$ state).}\label{fig:gc2q}
\end{figure*}

\begin{figure*}
\centering
\includegraphics[width=0.9\linewidth]{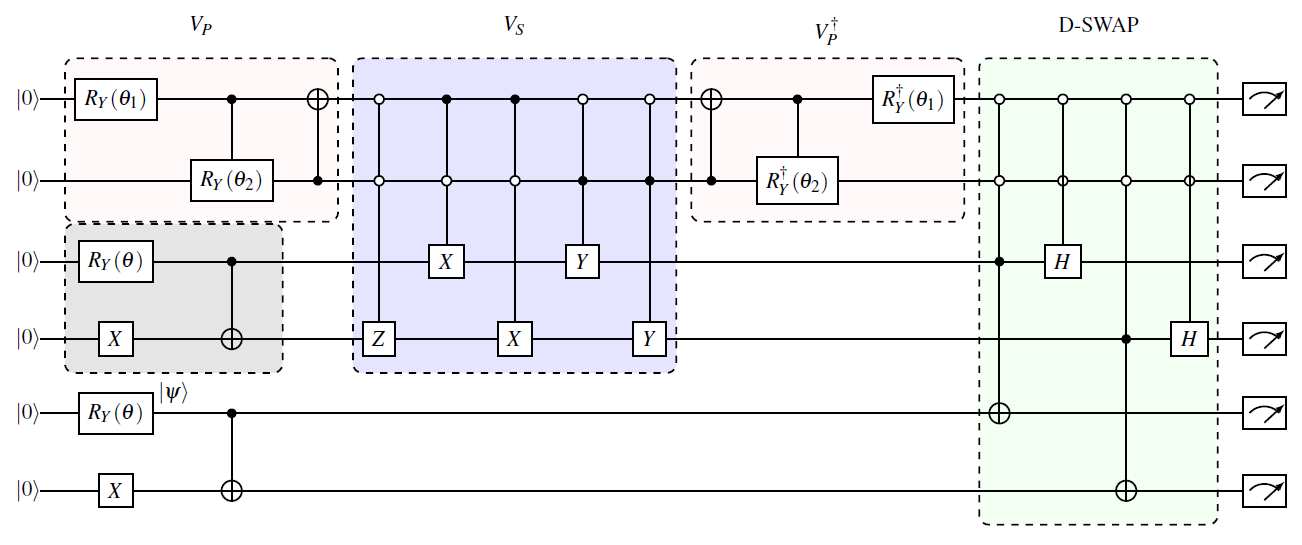}
\caption{Same as \figurename~\ref{fig:gc2q} but for Eq.~\eqref{eq:q2jw}, i.e., corresponding to JW transformation. Here $\theta_{1}=1.79286, \theta_{2}=1.5708$, and $\theta=0.559385$ (for the exact state obtained with classical calculations).}\label{fig:jw2q}
\end{figure*}

Based on LCU method, using the Eq.~\eqref{eq:vp1}, the \textit{prepare} operator $V_{P}$ can be written as
\begin{eqnarray}
    V_{P}\ket{0}&=&\sqrt{\frac{0.18144}{0.46538}}\ket{0}+\sqrt{\frac{0.28394}{0.46538}}\ket{1}\nonumber\\
    &=&R_{Y}(-3.700868)\ket{0},
\end{eqnarray}
and the \textit{selection} operator is given by
\begin{equation}
    V_{S}=\sum_{i=0}^{1}\ket{i}\bra{i}\otimes U_{i}=\ket{0}\bra{0}\otimes Z_{0} + \ket{1}\bra{1}\otimes X_{0}
\end{equation}
A complete circuit representing the above operations is given in \figurename~\ref{fig:gc2q}. Similar circuit for the JW case is given in \figurename~\ref{fig:jw2q}.

Though we study here only the one-body operators, the LCU based method can be used for two-body and other higher order operators as well. However, such operators will require more number of qubits and operations, making the simulations lesser fault tolerant.
%$$$$$$$$$$$$$$$$$$$$$$$

%$$$$$$$$$$$$$$$$$$$$$$$
\subsubsection{Using Hadamard test}
We can compute the expectation value of $\hat{Q}$ with Hadamard test (detailed in subsection~\ref{sec:htest}) for each unitary component separately.
\begin{figure*}%[h!]
\centering
\includegraphics[width=0.45\linewidth]{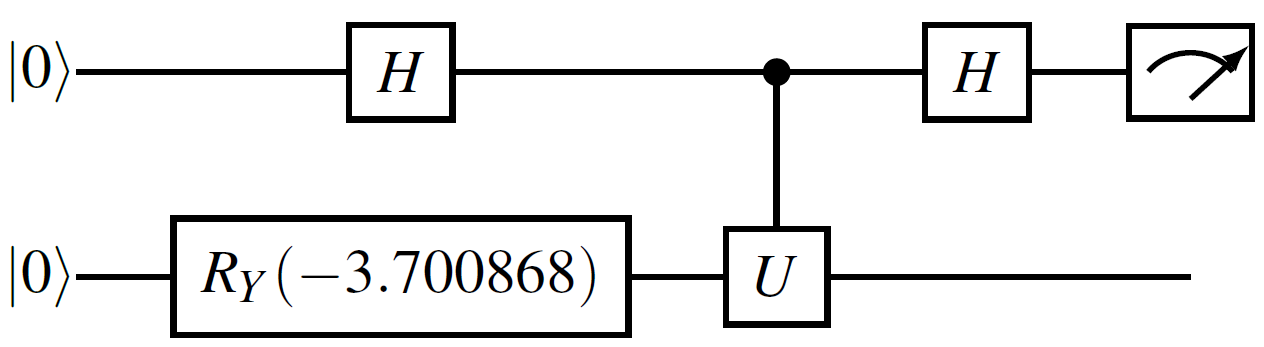}
\includegraphics[width=0.45\linewidth]{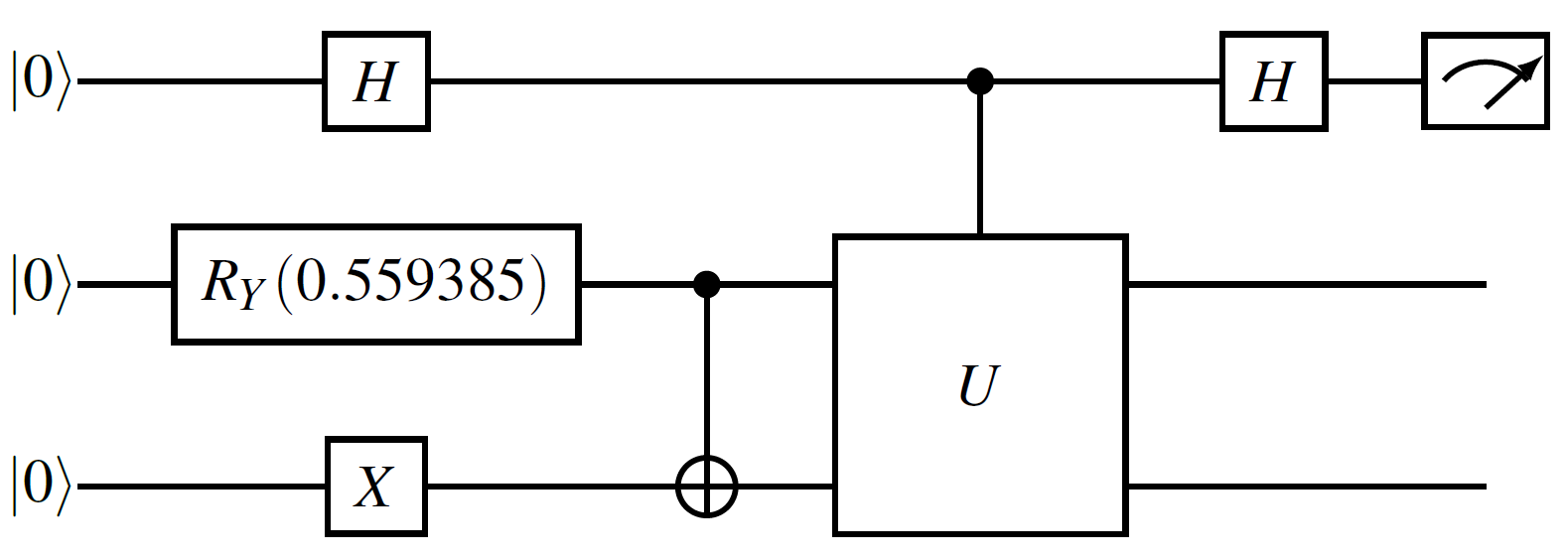}
\caption{The quantum circuits corresponding to Hadamard test (H-test) for the cases of GC (left) and JW (right) for calculating $\bra{\psi}U\ket{\psi}$ which is the difference between measuring the ancilla qubit in $0 (p(0))$ and $1 (p(1))$ states, i.e., $p(0)-p(1)$. For Eq.~\eqref{eq:q1} $U$ is $X_{0}$ and $Z_{0}$, and for Eq.~\eqref{eq:q2jw} $U$ is $Z_{1},X_{0}X_{1},$ and $Y_{0}Y_{1}$.}\label{fig:had}
\end{figure*}
To illustrate for the simplest case, we utilise the circuit shown in Fig.~\ref{fig:had} for the exact state calculated classically [Eq.~\eqref{eq:gc2psi}]. In case of one run of the circuit, for the $X_{0}$, we get $p(0)=0.76539$ and $p(1)=0.23461$, so $\bra{\psi}X_{0}\ket{\psi}=0.53078$, and for $Z_{0}$, we get $p(0)=0.0765$ and $p(1)=0.9235$ which gives $\bra{\psi}Z_{0}\ket{\psi}=-0.847$. Therefore, for the operator given in Eq.~\eqref{eq:q1}, we get $Q'=-0.18144\times0.847+0.28394\times0.53078=-0.00297$, and hence $Q=-0.18441$.

\section{Results}\label{sec:result}
First, we calculate the ground state energy and wave function for deuteron using the REID68 potential. We perform the calculations for basis sizes $2~(n=0;l=0,2)$ and $4~(n=0,1;l=0,2)$. We can obtain the results closer to experimental values with a basis size $n\rightarrow\infty$. However, the number of gates and operations required for such calculations is quite large. Hence, we stick to the simpler cases to understand the implementations of algorithms.
\begin{figure*}
    \centering
    \includegraphics[width=0.8\linewidth]{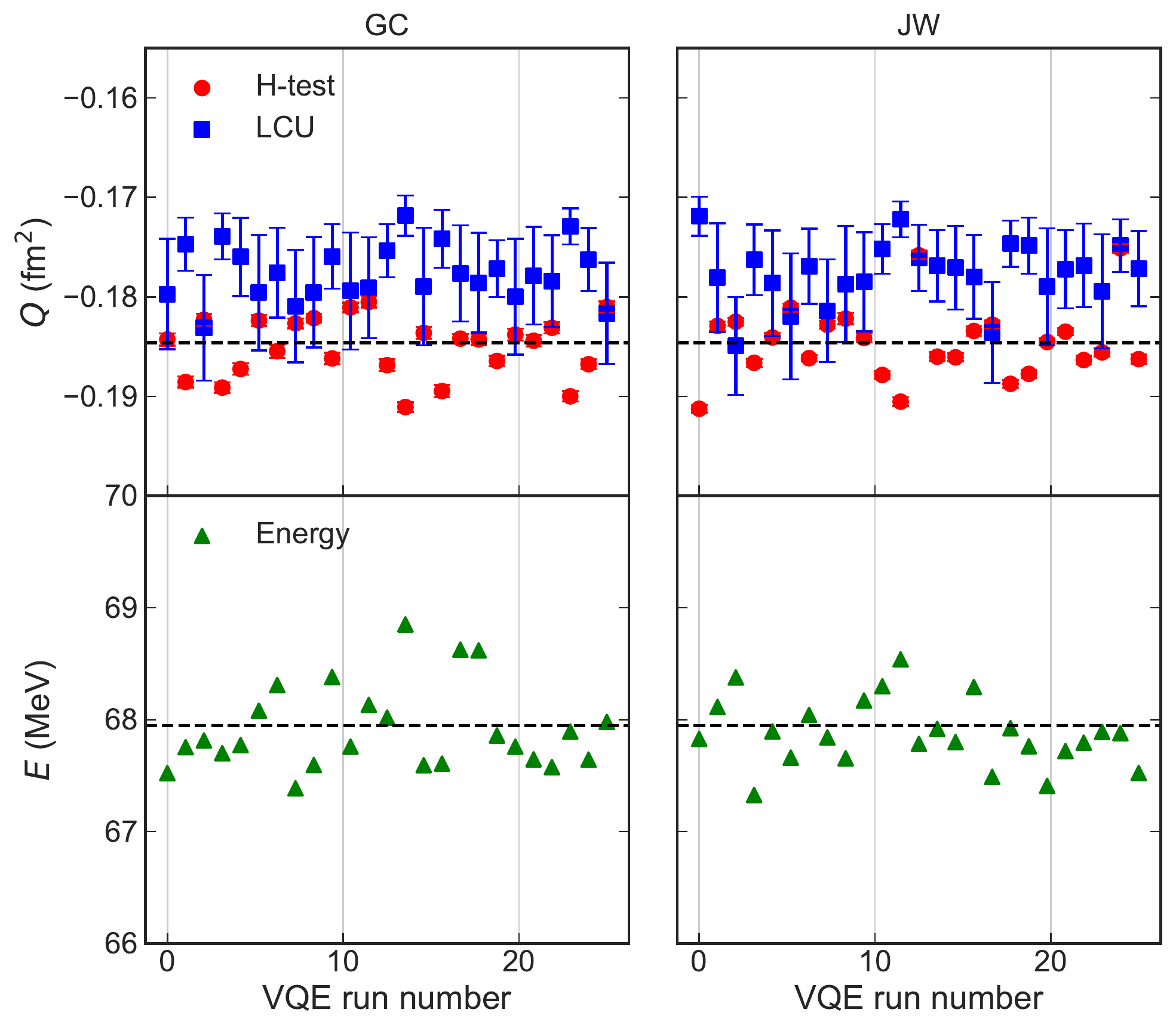}
    \caption{(a) The quadrupole moment for a basis size $2$ ($n=0;l=0,2$) obtained with the method based on H-test and LCU with GC (top left) and JW (top right), respectively, for a given run number of VQE with $100$ runs of LCU or H-test. The energies obtained using VQE with GC (bottom left) and JW (bottom right) are shown to compare the errors in quadrupole moment for each run due to the errors in wave function obtained with VQE. The black dashed lines represent the classically computed values.}
    \label{fig:2q}
\end{figure*}

\begin{figure*}
    \centering
    \includegraphics[width=0.8\linewidth]{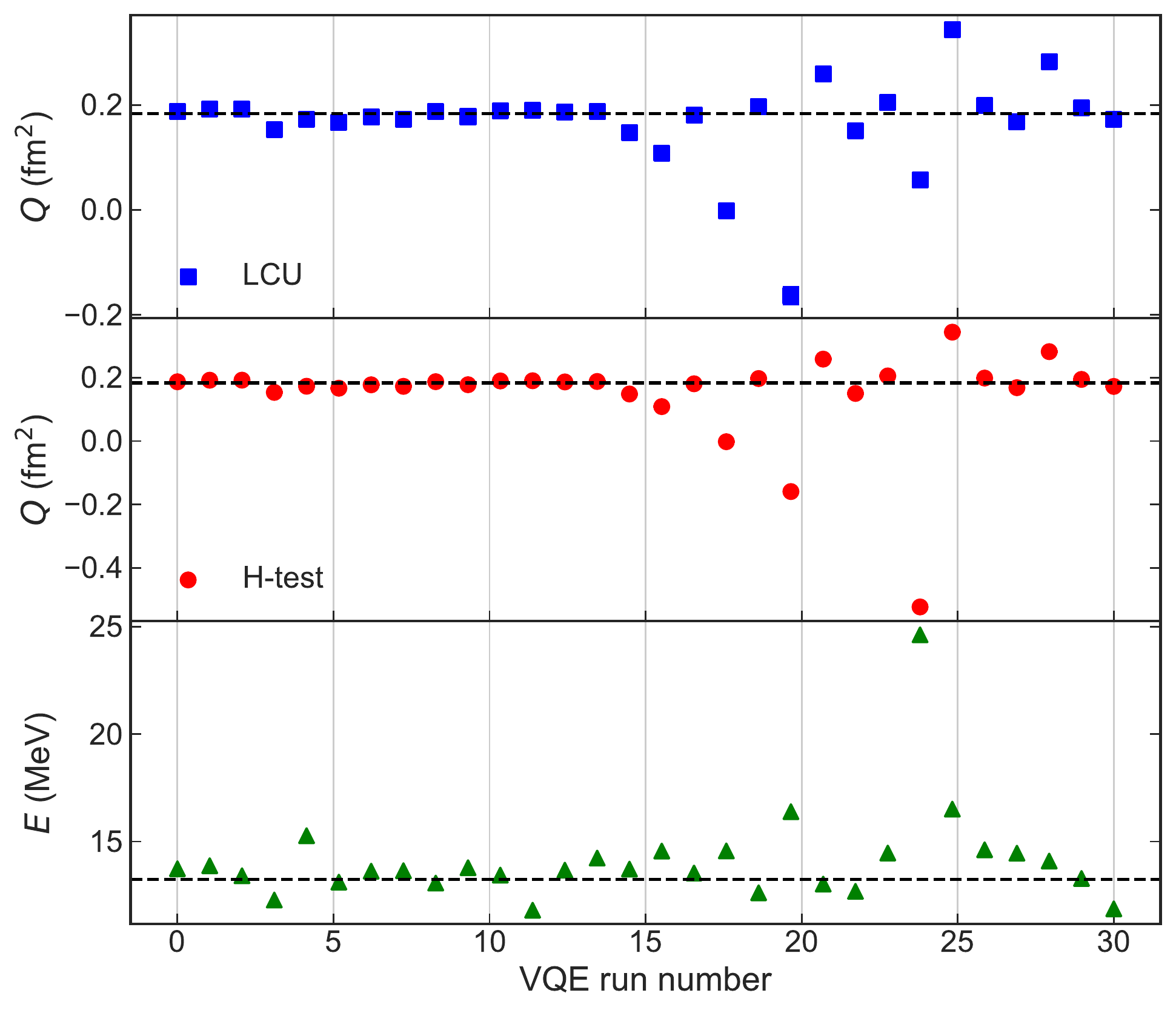}
    \caption{The quadrupole moment for a basis size $4$ ($n=0,1;l=0,2$) obtained with the method based on LCU (top) and H-test (center) along with the energy (bottom) obtained using VQE with GC for several runs.}
    \label{fig:4q}
\end{figure*}

We use IBM's Qiskit software package to perform the calculations with its QASM simulator. We execute $100$ independent runs of algorithms with several shots for each run. Number of shots required depends on the difference between $p(0)$ and $p(1)$ in the H-test, and the difference between probability of ancilla register being in state $0$ and others, to capture the correct value with precision. For example, if the expectation value of an operator is $0.01$, then, in case of H-test, the minimum number of shots should be $1000$. The estimation of typical value is given by median, and the median absolute deviation (MAD) is shown as the measure of errors. Furthermore, to explore the errors propagated due to the errors in VQE calculations, we execute the independent runs of VQE and utilise the obtained wave function from each run in the calculations of quadrupole moment. To compare the results, we show energies obtained in each run of VQE, and the corresponding quadrupole moment value calculated with LCU and H-test for basis sizes $2$ and $4$ in Figs.~\ref{fig:2q} and~\ref{fig:4q}, respectively.

The quadrupole moment is calculated with LCU and H-test methods and compared with the exact ones (calculated on conventional computer). As can be seen in \figurename~\ref{fig:2q}, the errors in calculated values are larger in case of LCU than H-test based methods. This behaviour lies in the fact that there are more number of gates and qubits required in LCU than in H-test. For example, in the case of GC, since the $\ket{\psi}$ is a single qubit state, H-test requires only two qubits (see Fig.~\ref{fig:had1}) for each term in the operator. Whereas, for the same case, LCU method requires four qubits with SWAP test and three qubits with D-SWAP test (see Fig.~\ref{fig:wcomp}). Consequently, multi-controlled gates (with control operations on more than one qubit unlike in H-test) appear in the circuits leading to more errors in simulations.

In case of basis size $4$, the error bars are not clearly visible due to the large scale as can be seen in \figurename~\ref{fig:4q}. This scale is a result of larger errors in VQE calculations due to more number of qubits and gates. Hence, as the wave function is farther away from the exact one, the errors are enhanced in quadrupole moment as well. However, the errors in LCU and H-test due to increasing number of qubits and gates (with increasing basis size) are enhanced only by a factor of two, leading to invisible error bars in Fig.~\ref{fig:4q}. This behaviour can be attributed to a smaller number of terms in the electric quadrupole moment operator as compared to the Hamiltonian. However, quantitatively, the results are better with the H-test than LCU based method.

In case of basis size $2$, we compare the results from GC and JW as shown in \figurename~\ref{fig:2q}. For a basis size $4$, the circuits become complicated in case of JW, and hence it is understood that errors will be larger in JW than in GC.

\begin{table*}%[h!]
    \centering
    \caption{Quadrupole moment for basis sizes $2$ and $4$ with different methods (H-test and LCU) and the energy in case of GC. The measure of errors is given by MAD.}
    \begin{tabular}{|c|c|c|c|c|c|c|c|}
    \hline
      Basis size  & \multicolumn{5}{|c|}{Quadrupole moment (fm$^{2}$)} & \multicolumn{2}{|c|}{Energy (MeV)}\\
    \cline{2-8}
      &Exact& \multicolumn{2}{|c|}{Without VQE} & \multicolumn{2}{|c|}{With VQE}  & Exact & VQE\\
      \cline{3-6}& & H-test & LCU & H-test & LCU & &\\
      \hline
       $2$  & $-0.1846$ & $-0.1846(5)$ & $-0.1785(16)$ & $-0.1848(24)$ &$-0.1805(37)$ & 67.948 & $67.848(233)$\\
       $4$  & $0.1836$ & ~$0.1838(10)$ & ~$0.1835(21)$ & ~$0.1784(173)$ & ~$0.1781(162)$ & 13.244 & $13.648(763)$\\
       \hline
    \end{tabular}
    \label{tab:1}
\end{table*}

To see the quantitative difference in errors for $Q$ with and without VQE, the explicit values are given in \tablename~\ref{tab:1}. For the case without VQE, we take the wave function as the classically calculated one. Supporting the previous arguments, the errors in quadrupole moment with the wave function from classical calculations (without VQE) are more in case of LCU than in case of H-test. Similar arguments are true with VQE, in case of basis size $2$, where we have calculated the median and MAD values utilising the results from $100$ runs of VQE with $100$ runs of LCU or H-test for each run of VQE (counting a total of $10000$ values for each case). However, in case of basis size $4$, the errors are large but almost equal with both the methods. Therefore, it would be interesting to explore how the errors in one step of the algorithm affect the errors in the next step.

\section{Conclusions}\label{sec:conc}
We employed the algorithms based on Hadamard test and linear combination of unitaries (LCU) to compute the expectation values of non-unitary operators on a quantum computer. We elaborated these techniques by implementing them to calculate the quadrupole moment of deuteron. The LCU based method is found to be more versatile as it can be used to calculate the overlap of excited state $\mathcal{O}\ket{\psi}$ with any other state different from $\ket{\psi}$. Whereas, in case of the Hadamard test based algorithm, one can calculate the overlap of $\mathcal{O}\ket{\psi}$ with $\ket{\psi}$ only. The Hadamard test based method is found to be more reliable as the errors are smaller due to lesser number of qubits and gates required as compared to the LCU based method. The Jordan-Wigner transformation and Gray code encoding are explored to compare the resource requirement and errors. The Gray code encoding is more efficient as the number of qubits are lesser. This becomes more crucial in the case of LCU based method. We further compared the results for exact wave functions calculated on a classical computer with the results obtained with the wave function calculated with variational quantum eigensolver (VQE) for several runs.  This work provides a premise to calculate several other observables that require to compute the expectation value of an operator (beyond the energies), through quantum simulations. It would be interesting to extend it to study other nuclear phenomena, for example, the resonances which involve the non-unitary operators.

\begin{acknowledgments}
This work was supported in part by the U.S.~Department of Energy, Office of Science, Office of High Energy Physics, under Awards No.~DE-SC0019465 and DE-FG02-95ER40907. 
\end{acknowledgments}
%$$$$$$$$$$$$$$$#######################$$$$$$$$$$$$$$$$$$$$$$$$$$$$$$$$$$$$$$$
%$$$$$$$$$$$$$$$#########   Appendices   ######$$$$$$$$$$$$$$$$$$$$$$$$$$$$$$$
%$$$$$$$$$$$$$$$#######################$$$$$$$$$$$$$$$$$$$$$$$$$$$$$$$$$$$$$$$
\appendix
\section{SWAP and destructive SWAP test}\label{app:swap}

% \begin{figure}[h!]
%      \centering
%      \begin{subfigure}%{\textwidth}
%          \centering
%          \includegraphics[width=0.7\linewidth]{swap1.png}
%          \caption*{$(a)$}
%      \end{subfigure}
%      \begin{subfigure}%{\textwidth}
%          \centering
%          \includegraphics[width=0.7\linewidth]{swap2.png}
%          \caption*{$(b)$}
%      \end{subfigure}
%      \begin{subfigure}%{\textwidth}
%          \centering
%          \includegraphics[width=\linewidth]{swap3.png}
%          \caption*{$(c)$}
%      \end{subfigure}
%      \begin{subfigure}%{\textwidth}
%          \centering
%          \includegraphics[width=\linewidth]{swap4.png}
%          \caption*{$(d)$}
%      \end{subfigure}
%      \begin{subfigure}%{\textwidth}
%          \centering
%          \includegraphics[width=0.7\linewidth]{swap5.png}
%          \caption*{$(e)$}
%      \end{subfigure}
%         \caption{Reducing the quantum circuit (a) corresponding to SWAP test~\cite{swap:2013} to an efficient one (e).}
%     \label{fig:swap}
% \end{figure}
% \begin{figure}[h!]
%     \centering
% \includegraphics[width=\linewidth]{2q_swap.png}
%     \caption{Extending the quantum circuit for SWAP test to multiple qubits is straightforward as shown here in the case of $2$-qubits.}
%     \label{fig:swap2q}
% \end{figure}

\begin{figure}[h!]
     \centering
         \includegraphics[width=0.7\linewidth]{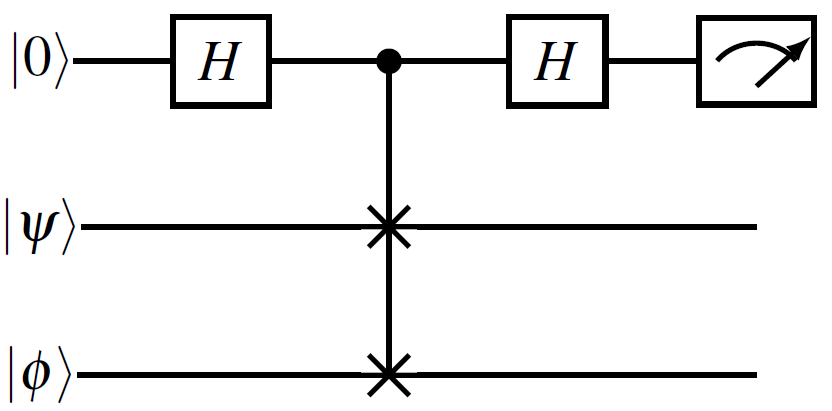}
         \caption*{$(a)$}
         \includegraphics[width=0.7\linewidth]{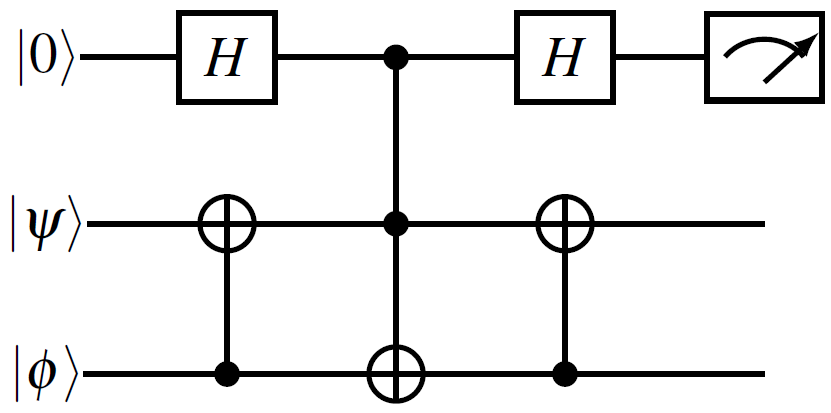}
         \caption*{$(b)$}
         \includegraphics[width=\linewidth]{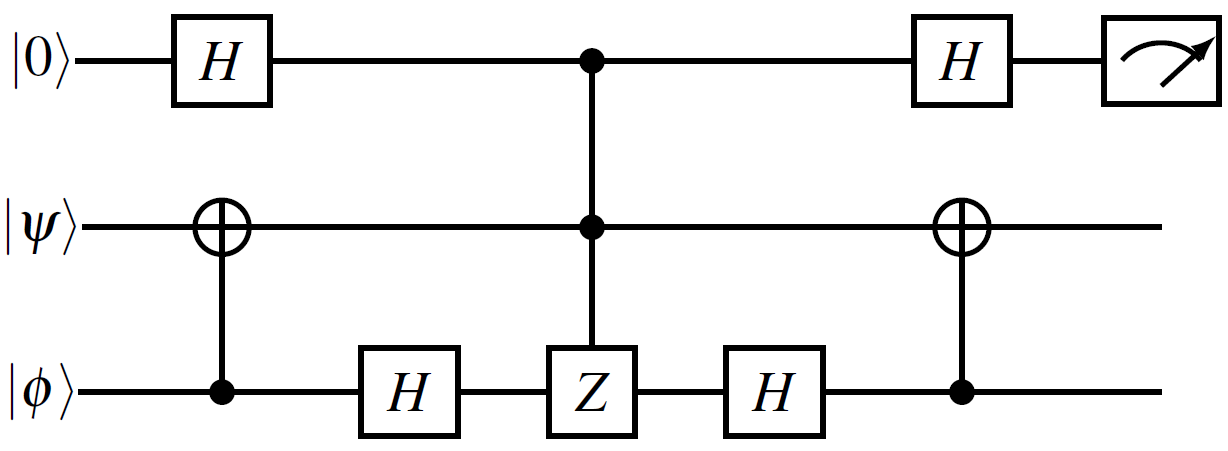}
         \caption*{$(c)$}
         \includegraphics[width=\linewidth]{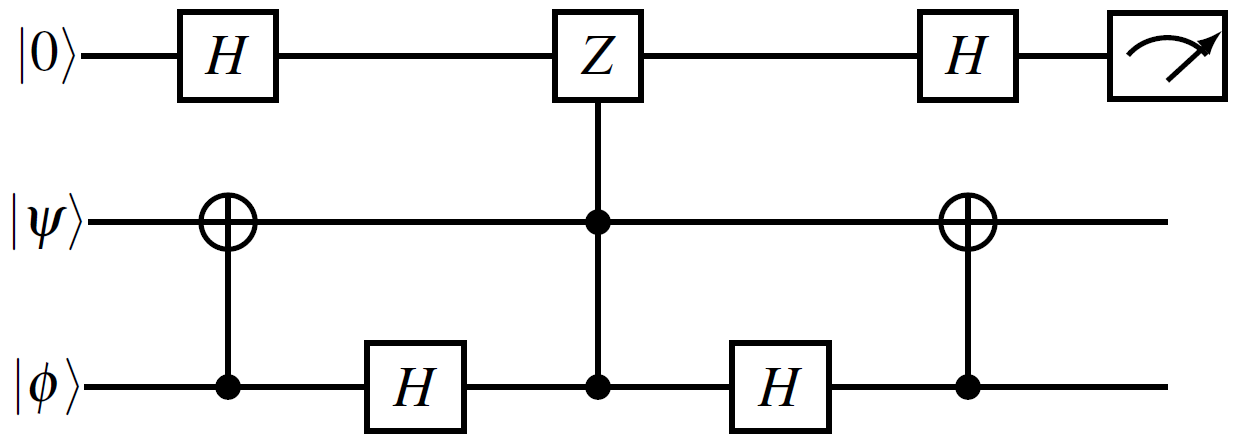}
         \caption*{$(d)$}
         \includegraphics[width=0.7\linewidth]{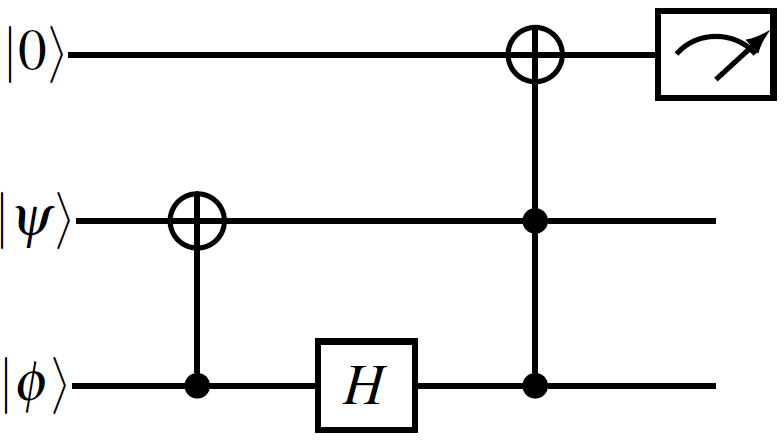}
         \caption*{$(e)$}
        \caption{Reducing the quantum circuit (a) corresponding to SWAP test~\cite{swap:2013} to an efficient one (e).}
    \label{fig:swap}
\end{figure}
\begin{figure}[h!]
    \centering
\includegraphics[width=\linewidth]{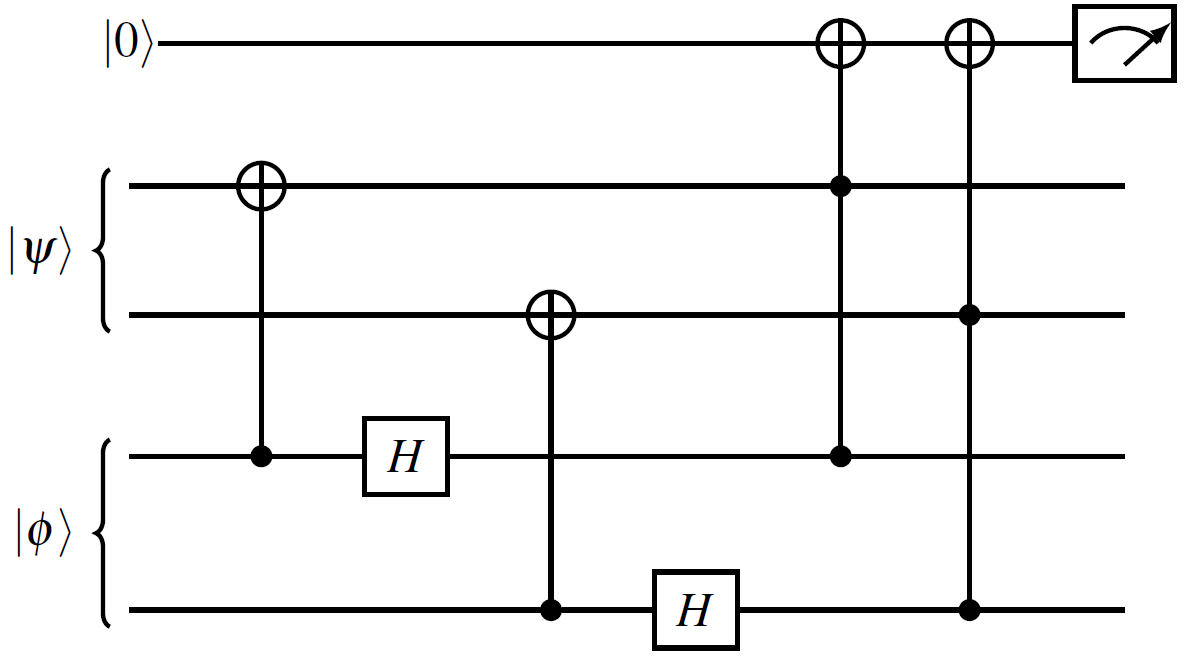}
    \caption{Extending the quantum circuit for SWAP test to multiple qubits is straightforward as shown here in the case of $2$-qubits.}
    \label{fig:swap2q}
\end{figure}
%$$$$$$$$$$$$$$$$$$$$$$$$$$$$$$$$$
\begin{figure}[h!]
    \centering
\includegraphics[width=0.45\linewidth, valign=c]{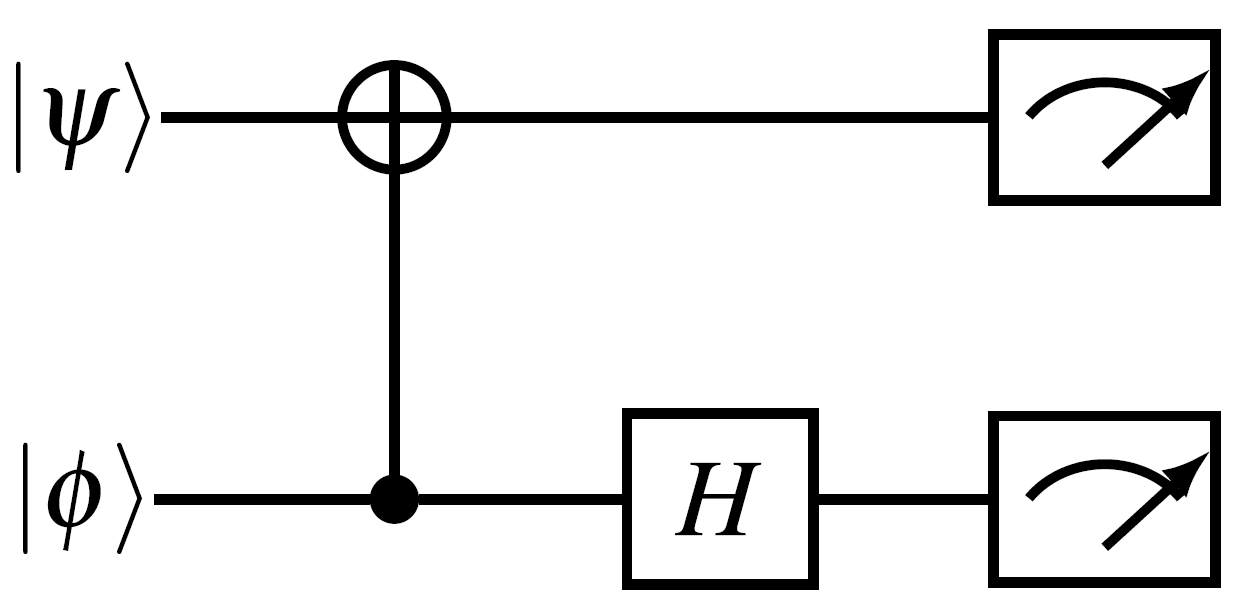}
% \qquad
$\equiv$
% \qquad
\includegraphics[width=0.45\linewidth, valign=c]{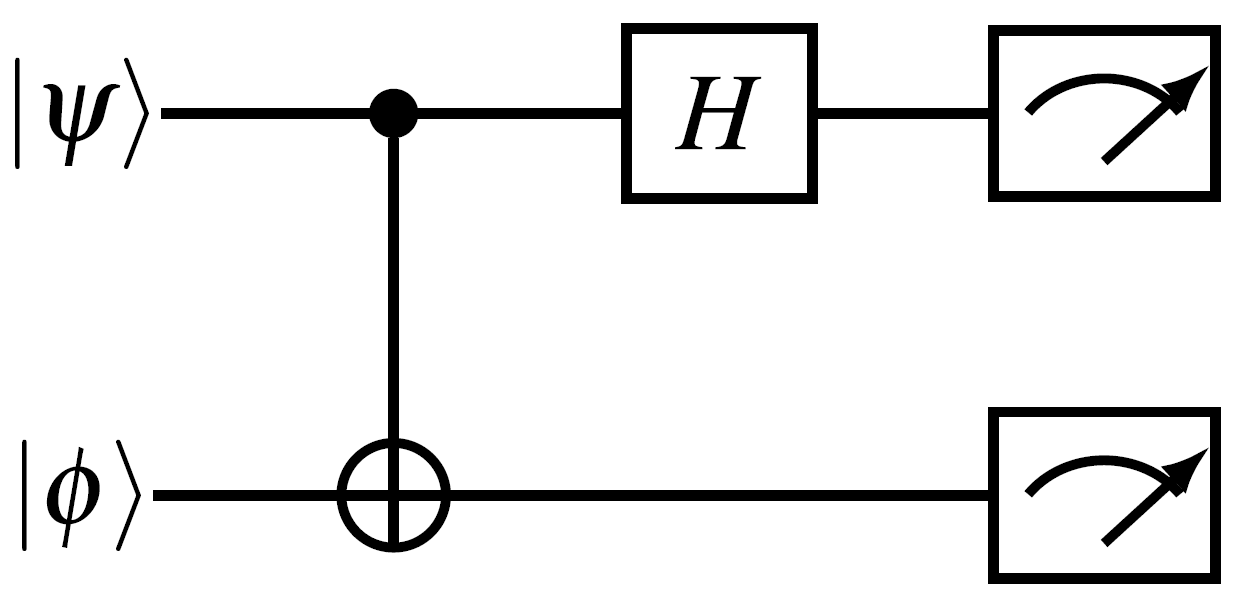}
    \caption{The quantum circuit corresponding to destructive SWAP test.}
    \label{fig:dswap}
\end{figure}
The SWAP test which could reveal the overlap between two states is discussed with more details in Ref.~\cite{swap:2013}. Here, we provide the crucial details, starting with the corresponding circuits in \figurename~\ref{fig:swap}. The SWAP operation consists of three alternate CNOT gates which interchange the input states $\ket{\psi}$ and $\ket{\phi}$. To test the overlap of input states, we require an ancillary qubit on which a Hadamard gate $H$ is applied before and after the controlled SWAP operation on $\ket{\psi}$ and $\ket{\phi}$. Measuring the ancillary qubit provides us the overlap $|\braket{\psi|\phi}|^{2}$. After the SWAP test, the input states are impossible to recover due to the entanglement.

This circuit can be further simplified as shown in \figurename~\ref{fig:swap} (b-e). In \figurename~\ref{fig:swap} (b), if ancillary qubit is $0$, the middle CNOT gate has no effect and hence left and right CNOT gates cancel each others effect. When ancillary qubit is $1$, it works as a usual SWAP gate. Furhtermore, utilizing the facts $X=HZH$ and $CX=H(CZ)H$, we get \figurename~\ref{fig:swap} (c). Noticing that the ancillary qubit is not affected after the controlled $Z$ gate, we can ignore the CNOT and $H$ gates after $CCZ$. The target qubit has been changed in \figurename~\ref{fig:swap} (d) based on the fact that a sign change occurs only when all three qubits are $1$. In \figurename~\ref{fig:swap} (d), $HZH$ is replaced by $X$. The circuit for SWAP test for two qubit states is given in \figurename~\ref{fig:swap2q} which can be extended for multiple qubit states similarly.

We note that ancillary qubit changes only when both states $\ket{\psi}$ and $\ket{\phi}$ are $1$ after CNOT and $H$ operations. With this prior knowledge, we do not need an ancillary qubit. Instead, we can measure the input states after CNOT and $H$ operations. The resulting ovelap $|\braket{\psi|\phi}|^{2}$ can be interpreted as the difference between the probability of both input states being $1$ and the rest. This protocol is known as destructive SWAP test~\cite{swap:2013}. The corresponding circuit is given in \figurename~\ref{fig:dswap}.
\section{Preparation of $V_{p}\ket{0}^{\otimes n_{a}}$}\label{app:vp}
If an operator written in terms of linear combination of unitaries $U$s is given by
\begin{equation}
    \hat{\mathcal{O}}=\sum_{i=0}^{k-1}\beta_{i}U_{i}
\end{equation}
where $k$ is the total number of terms in the operator. With $\Lambda=\sum_{i}|\beta_{i}|$, we prepare the circuit $V_{p}$ which leads to the following state
\begin{eqnarray}
V_{p}\ket{0}^{\otimes n_{a}}=\ket{\phi}&=&\sum_{i}a_{i}\ket{i}.
\end{eqnarray}
where $a_{i}=\sqrt{\frac{\beta_{i}}{\Lambda}}$. For convenience, we take all the $\ket{i}$s in binary representation, for example, $\ket{3}\equiv\ket{0_{n_{a}}\ldots0_{2}1_{1}1_{0}}$. With increase in $i$, only one qubit changes from $0$ to $1$. As we will see in the forthcoming discussion, it leads to a simpler circuit designing.

We start from $\ket{i}=\ket{0}$ and add $\ket{i}$th state with corresponding coefficient $a_{i}$ in increasing order. Moving from $\ket{i-1}$ to $\ket{i}$ we apply the $R_{Y}(2\theta_{i})$ on the qubit that changes from $0$ to $1$ with the control operation on the qubits which are $1$ in $\ket{i-1}$. Next, we apply CNOT gates on the qubits those change from $1$ in $\ket{i-1}$ to $0$ in $\ket{i}$ with control on the qubits which are $1$ in $\ket{i}$. In this scheme we need $k-1$ rotation gates with $\theta_{i}=\cos^{-1}\left(\frac{a_{i}}{\sqrt{1-\sum_{j<i}a_{j}^{2}}}\right)$.

\begin{figure*}%[h!]
    \centering
\includegraphics[width=0.8\linewidth]{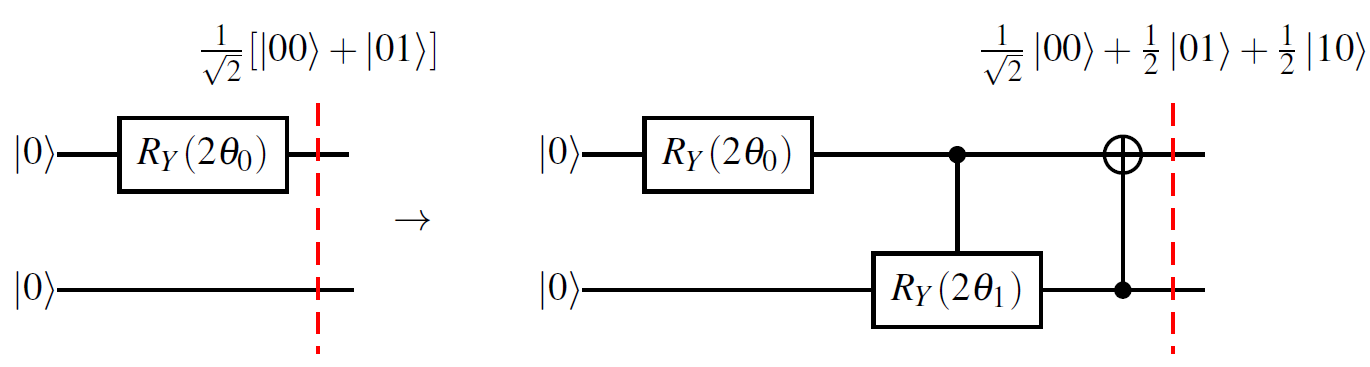}
    \caption{The quantum circuit corresponding to preparation of $\ket{\phi}$ given in Eq.~\eqref{eq:vp}.}
    \label{fig:vp}
\end{figure*}

To see an example, let us consider an operator given by
\begin{equation}
    \hat{\mathcal{O}}=2X_{0}+X_{0}Y_{1}+Z_{1}.
\end{equation}
In this case, we have $\Lambda=4$ and we need to prepare the state
\begin{eqnarray}\label{eq:vp}
    \ket{\phi}&=&\frac{1}{\sqrt{2}}\ket{0} + \frac{1}{2}\ket{1}+\frac{1}{2}\ket{2}\nonumber\\
    &\equiv&\frac{1}{\sqrt{2}}\ket{00} + \frac{1}{2}\ket{01}+\frac{1}{2}\ket{10}.
\end{eqnarray}
We need $\left\lceil\log_{2}3\right\rceil=2$ qubits to implement the corresponding quantum circuit. First we prepare $\ket{0}$ by initializing all qubits to $0$. To prepare a state with $\ket{0}$ and $\ket{1}$, since first qubit changes from $0$ to $1$, we apply a $R_{Y}(2\theta_{0})$ gate on first qubit with $\theta_{0}=\cos^{-1}\left(\frac{1}{\sqrt{2}}\right)$. The resulting circuit is shown in \figurename~\ref{fig:vp}(a). Moving from $\ket{1}$ to $\ket{2}$, second qubit changes from $0$ to $1$, we apply $R_{Y}(2\theta_{1})$ gate on second qubit with $\theta_{1}=\cos^{-1}\left(\frac{1}{\sqrt{2}}\right)$ controlled by first qubit (since it is $1$ in $\ket{1}$). Furthermore, as the first qubit changes from $1$ to $0$, we apply CNOT gate on the first qubit controlled by second the qubit (since it is $1$ in $\ket{2}$). These steps are illustrated in \figurename~\ref{fig:vp}. This circuit can be further simplified by utilising the techniques given in Ref.~\cite{nielsen2002}.

% $$$$$$$$$$$$$$$$$$$$$ Bibliography $$$$$$$$$$$$$$$$$

\bibliographystyle{apsrev} 
\bibliography{ref}

\end{document}